\documentclass[prd,twocolumn,superscriptaddress,showpacs,amssymb,amsmath,amsfonts,aps,altaffilletter,nofootinbib]{revtex4}

\usepackage[british]{babel}
\usepackage{float}
\floatstyle{boxed}

\usepackage{graphicx,bm,dcolumn,amsmath,amsfonts,subfigure,color,latexsym,amssymb}

\def\be{\begin{equation}}
\def\ee{\end{equation}}

\def\bea{\begin{eqnarray}}
\def\eea{\end{eqnarray}}

\def\nn{\nonumber}

\begin{document}

\title[Template banks for spinning black hole binaries]
{Template banks to search for compact binaries with spinning
components in gravitational wave data }

\author{Chris Van Den Broeck}
\email{Chris.van-den-Broeck@astro.cf.ac.uk}
\affiliation{School of Physics and Astronomy, Cardiff University, 
5 The Parade, Cardiff, CF24 3YB, UK}
\author{Duncan A. Brown}
\email{dabrown@phys.syr.edu}
\affiliation{Department of Physics, Syracuse University,
Syracuse, NY 13244, USA}
\author{Thomas Cokelaer}
\email{Thomas.Cokelaer@astro.cf.ac.uk, Thomas.Cokelaer@inria.fr}
\affiliation{School of Physics and Astronomy, Cardiff University, 
5 The Parade, Cardiff, CF24 3YB, UK}
\author{Ian Harry}
\email{Ian.Harry@astro.cf.ac.uk}
\affiliation{School of Physics and Astronomy, Cardiff University, 
5 The Parade, Cardiff, CF24 3YB, UK}
\author{Gareth Jones}
\email{Gareth.Jones@astro.cf.ac.uk}
\affiliation{School of Physics and Astronomy, Cardiff University, 
5 The Parade, Cardiff, CF24 3YB, UK}
\author{B.~S.~Sathyaprakash} 
\email{B.Sathyaprakash@astro.cf.ac.uk} 
\affiliation{School of Physics and Astronomy, Cardiff University, 
5 The Parade, Cardiff, CF24 3YB, UK}
\author{Hideyuki Tagoshi}
\email{tagoshi@vega.ess.sci.osaka-u.ac.jp}
\affiliation{Department of Earth and Space Science, 
Graduate School of Science, Osaka University, Toyonaka 560-0043, Japan}
\author{Hirotaka Takahashi}
\email{hirotaka@oberon.nagaokaut.ac.jp}
\affiliation{Department of Management and Information Systems Science,
Nagaoka University of Technology, Nagaoka, Niigata 940-2188, Japan}

\begin{abstract}
Gravitational waves from coalescing compact binaries are one of the most
promising sources for detectors such as LIGO, Virgo and GEO600. If the
components of the binary posess significant angular momentum (spin), as is
likely to be the case if one component is a black hole, spin-induced
precession of a binary's orbital plane causes modulation of the
gravitational-wave amplitude and phase. If the templates used in a
matched-filter search do not accurately model these effects then the sensitivity,
and hence the detection rate, will be reduced. We investigate the ability of
several search pipelines to detect gravitational waves from compact binaries
with spin. We use the post-Newtonian approximation to model the inspiral phase
of the signal and construct two new template banks using the phenomenological
waveforms of Buonanno, Chen and Vallisneri [A.~Buonanno, Y.~Chen and M.~Vallisneri, Phys.~Rev.~D67, 104025 (2003)]. We compare
the performance of these template banks to that of banks constructed using the
stationary phase approximation to the non-spinning post-Newtonian inspiral
waveform currently used by LIGO and Virgo in the search for compact binary
coalescence. We find that, at the same false alarm rate, a search pipeline
using phenomenological templates is no more effective than a pipeline which
uses non-spinning templates. We recommend the continued use of the
non-spinning stationary phase template bank until the false alarm rate
associated with templates which include spin effects can be 
substantially reduced.
\end{abstract}

\maketitle
\date{today}
\section{Introduction}
\label{s:intro}

In 2005-2007, the Laser Interferometer Gravitational-Wave Observatory (LIGO)  
recorded two years of data at design sensitivity~\cite{Abbott:2007kva} and
the LIGO, Virgo~\cite{Acernese:2005yh} and GEO600~\cite{Willke:2007zz} detectors now
form a world-wide network of broad-band gravitational-wave observatories. The 
LIGO and Virgo detectors are scheduled to resume operations in
summer 2009 with a factor of $\sim 2$--$3$ sensitivity increase over
previous observations. 
The gravitational waves emitted during the inspiral and merger of binaries
containing neutron stars (NS) and/or black holes (BH) are a primary target of
this network.
Binary neutron stars (BNS) can be observed up to 35~Mpc (70~Mpc) in the
Initial (Enhanced) LIGO detectors and up to 450~Mpc in the Advanced LIGO
detectors, which will begin observations in 2015~\cite{Abbott:rates}.
Binary black holes (BBH) with $10 M_\odot$ components 
should be visible at $170$~Mpc (350~Mpc) in the Initial (Enhanced) LIGO
detectors, increasing to 2 Gpc in Advanced LIGO~\cite{Abbott:rates}.
Population synthesis calculations constrained by radio observations of BNS
systems containing pulsars predict BNS detection rates between $10^{-3}$--$1\ \text{yr}^{-1}$ for
Enhanced LIGO and $0.4$--$400 \ \text{yr}^{-1}$ for Advanced
LIGO, with the most likely values being $0.1\ \text{yr}^{-1}$ and $40\
\text{yr}^{-1}$, respectively~\cite{Kopparapu:2007ib,Kalogera:2003tn,Abbott:rates}. Much less is
known about the detection rates of BBH and NS-BH coalescences, although it is
plausible that Enhanced (Advanced) LIGO rates could be as high as 
$1\ (300) \ \text{yr}^{-1}$ for NS-BH binaries and $20\ (4000) \ \text{yr}^{-1}$
for BBH \cite{O'Shaughnessy:2006wh,O'Shaughnessy:2005qs,Abbott:rates}.


The sensitivities listed in the preceding paragraph are optimal: they assume  accurate knowledge of the signal
waveform in order to construct matched filters which can extract
gravitational-wave signals buried in the noisy detector
data~\cite{Wainstein:1962,thorne.k:1987}.  The gravitational waveform from 
the inspiral of two compact objects has been calculated using the
post-Newtonian (PN) approximation, which uses the characteristic velocity of
the binary $(v/c)$ as an expansion
parameter~\cite{Blanchet:1995fg,Will:1996zj,Blanchet:1996pi,Blanchet:1996wx,Blanchet:1995ez,Blanchet:2001ax,Blanchet:2005tk,Blanchet:2004ek,Arun:2004ff,Blanchet:2008je,Kidder:1992fr,Kidder:1995zr,Faye:2006gx,Blanchet:2006gy,Arun:2008kb}.
Ongoing comparisons of PN waveforms with numerical simulations of binary black
holes have thus far confirmed the accuracy of the PN
solution in the late stages of inspiral~\cite{Hannam:2007ik,Buonanno:2006ui,Boyle:2008ge}, although optimal
searches for sources with total mass $\gtrsim 30\ M_\odot$ in the
first-generation detectors require waveforms that also model the merger and
ringdown~\cite{Buonanno:2007pf,Buonanno:2009qa}.  If the components of the
binary have negligible instrinsic angular momentum (spin) then it is straightforward to
construct a \emph{bank} of matched filters, parameterized by the two component
masses of the binary, and use these filters to search for
signals~\cite{Owen:1995tm,Owen:1998dk,Babak:2006ty,Allen:2005fk}. However, if
the components of a binary are spinning, then these spins can couple with the
orbital angular momentum of the binary and with each other to cause amplitude
and phase modulation of the gravitational waveform~\cite{Apostolatos:1994mx}.
Attempting to detect gravitational waves from spinning binaries with
non-spinning templates will result in a sub-optimal search and a corresponding
reduction in the detection rate~\cite{Apostolatos:1995pj,Apostolatos:1996rg}.
Since it is possible that a large fraction of astrophysical black holes 
have considerable spin~\cite{Kalogera:1999tq,O'Shaughnessy:2005qc}, it is
important to consider the effect of spin in searches for gravitational waves
from BBH and NS-BH coalescences. However, optimal
filters for spinning binaries are characterized by a much larger number of parameters than the ones for
non-spinning binaries, complicating placement of filters in the bank and
considerably increasing the computational cost of searches.

To mitigate the computational problem without compromising the sensitivity
of the search, a phenomenological family of templates was proposed by
Buonanno, Chen and Vallisneri~\cite{Buonanno:2002fy} (we refer to these
templates as BCV-spin). Filters constructed from these templates are described
by only four parameters and have good overlaps with the full PN
waveforms~\cite{Buonanno:2002fy}. Moreover, constructing a bank of filters
using BCV-spin waveforms is straightforward, if
cumbersome~\cite{Buonanno:2005pt,Abbott:2007ai}. The first searches for binary
black hole signals in LIGO data used non-spinning
templates~\cite{Abbott:2005kq,Abbott:2007xi}, however BCV-spin templates were 
recently used to search for BBH and NS-BH signals with spin in data from the
third LIGO science run~\cite{Abbott:2007ai}. The sensitivity of the search
described in~\cite{Abbott:2007ai} was not as good as the main results of
Ref.~\cite{Buonanno:2002fy} might suggest. This was primarily due to the the
response of the BCV-spin template to the non-gaussian noise transients  present
in real gravitational-wave detector data and the increase in the number of
degrees of freedom associated with the detection statistic (due to the larger
search parameter space)~\cite{Abbott:2007ai}. This was already anticipated in \cite{Buonanno:2002fy}; here we provide a quantitative analysis.

In this paper, we present an improvement to the search pipeline described
in~\cite{Abbott:2007ai}, by constructing banks that are much better suited to the BCV-spin template family. We compare the sensitivity of this search to the search for gravitational waves from compact binary coalescence with non-spinning templates in LIGO data~\cite{Babak:2006ty,Allen:2005fk}. Our
main conclusion is that while the BCV-spin templates have rather good overlaps
with the target waveforms, the current search pipeline needs further
improvements before any gains from these increased overlaps can be realized.  The false
alarm rate of BCV-spin filters in real detector data is larger than that of a
non-spinning search. This makes a search using BCV-spin templates less
sensitive than a non-spinning search when looking for binaries with spin,
since one has to use a higher detection threshold to obtain the same false
alarm rate. The results of this work were used to guide the decision \emph{not} to
implement the BCV-spin search on data from the fifth LIGO science
run and instead to use non-spinning filters to search for binaries with
spin~\cite{Abbott:2009tt}. The motivation for this decision was summarized in an appendix of
Ref.~\cite{Abbott:2009tt} and this paper can be seen as a companion to that
work. Here we present a detailed account of how the BCV-spin banks
were constructed, and how the comparisons between the BCV-spin and
non-spinning searches were performed.

This paper is organized as follows.  In Sec.~\ref{s:TargetWaveforms} we give a
description of our target signals, which are post-Newtonian waveform models for signals from spinning black hole
binaries, followed by a summary, in Sec.~\ref{s:DTF}, of the phenomenological BCV-spin
templates of Ref.~\cite{Buonanno:2002fy}. In Sec.~\ref{s:Banks} we review the
construction of the template bank used in Ref.~\cite{Abbott:2007ai} and
present two new methods to construct BCV-spin template banks, relaxing the
assumptions used in Ref.~\cite{Abbott:2007ai}: a ``square-hexagonal''
placement which generalizes the hexagonal placement developed
in~\cite{Cokelaer:2007kx} to a higher-dimensional template manifold, and a
stochastic placement technique proposed in~\cite{Allen:2009}.  In
Sec.~\ref{s:Matches} we compute matches of these banks with target waveforms
and compare the results with those obtained from two-dimensional template
banks based on the stationary phase approximation
(SPA)~\cite{Sathyaprakash:1991mt,Droz:1999qx} and hexagonal template placement
used in LIGO's non-spinning
searches~\cite{Abbott:2003pj,Abbott:2005pe,Abbott:2005pf,Abbott:2005kq,Abbott:2005qm,Abbott:2007xi,Abbott:2009tt}.
We compare the detection efficiency of the spinning banks with that of SPA
banks in Sec.~\ref{s:False alarms} followed by concluding remarks in
Sec.~\ref{s:Conclusions}.  Throughout this paper, we set $G=c=1$ unless
otherwise stated.


\section{Post-Newtonian waveforms from spinning binaries}
\label{s:TargetWaveforms}

Depending on their birth spins,
BHs in binaries could accumulate significant spin through accretion~\cite{Kalogera:1999tq,Belczynski:2008}. There is much
uncertainty concerning the equation of state of a neutron star, but most
models place an upper limit $J/M^2 \lesssim 0.7$ on the spin, above which the
star would break up~\cite{Cook:1993qr}. There is also an upper limit for the
spin of a black hole due to torque caused by radiation from the accretion disk
getting swallowed by the BH, leading to an expected bound of $J/M^2 \lesssim
0.998$~\cite{Thorne:1974ve}. Most of the modeling of spin evolution in compact
binaries has been confined to NS-BH systems, in which case the spin tilt with
respect to the orbital angular momentum can be
considerable~\cite{Kalogera:1999tq,Belczynski:2008}; this may also be the case for BBHs.

PN theory has achieved great success in modeling the adiabatic, quasi-circular
phase of inspiral, during which the fractional change in the orbital frequency
over each orbital period will be negligible (see, e.g.,
Ref.~\cite{Blanchet:2002av} for a review).  The orbital phasing has been
calculated to order $(v/c)^7$ (or 3.5PN in the usual
notation)~\cite{Damour:2000zb,
Blanchet:1995ez,Blanchet:1995fg,Will:1996zj,Blanchet:1996pi,Blanchet:1996wx,Blanchet:2001ax,Blanchet:2001aw,Damour:2000we,Blanchet:2004ek,Blanchet:2003gy,Blanchet:2004bb,Blanchet:2004re,Blanchet:2005tk} 
while the gravitational-wave amplitude for non-spinning binaries has been calculated
to order $(v/c)^6$~\cite{Blanchet:1996pi,Arun:2004ff,Blanchet:2008je}. The effect of spin on the
gravitational-wave phasing is known to order
$(v/c)^5,$~\cite{Kidder:1992fr,Kidder:1995zr,Faye:2006gx,Blanchet:2006gy} and
to order $(v/c)^3$ for the amplitude~\cite{Arun:2008kb}. However, 
since the matched filter is most sensitive to the phase evolution of the
binary, template waveforms amplitudes are typically computed only at leading
order in amplitude (the restricted waveform). Spin-orbit
interaction enters the phasing at 1.5PN and 2PN order and spin-spin
interaction at 2PN order. Spin effects influence the evolution of the orbital
frequency as a function of time. Including these effects, the adiabatic
evolution of the orbital frequency $\omega(t)$ is given by\footnote{At the
time this work was started the spin-orbit term at
2.5PN~\cite{Faye:2006gx,Blanchet:2006gy} was not yet known and so is not
included here.}
\bea
\frac{\dot{\omega}}{\omega^2} &=& \frac{96}{5}\eta(M\omega)^{5/3}\left[
1-\frac{743+924\eta}{336}(M\omega)^{2/3}\right.\nonumber\\ 
& - & \left \{ \sum_{i=1}^2 \frac{\chi_i(\mathbf{\hat{L}}_N\cdot\mathbf{\hat{S}}_i)}{12}\left(113\frac{m_i^2}{M^2}+75\eta\right) - 4\pi\right \} M\omega\nonumber\\
& + & \left(\frac{34103}{18144} + \frac{13661}{2016}\eta + \frac{59}{18}\eta^2 - \frac{\eta\chi_1\chi_2}{48} \left [247(\mathbf{\hat{S}}_1\cdot\mathbf{\hat{S}}_2) \right. \right. \nonumber \\
 & - & \left. \left. 721(\mathbf{\hat{L}}_N\cdot\mathbf{\hat{S}}_1)(\mathbf{\hat{L}}_N\cdot\mathbf{\hat{S}}_2) \right ] \right ) (M\omega)^{4/3}
\nonumber\\ 
& - &\frac{1}{672}(4159+15876\eta)\pi(M\omega)^{5/3}
\nonumber\\
& + &\left\{\left(\frac{16447322263}{139708800} - \frac{1712}{105}\gamma + \frac{16}{3}\pi^2\right) 
\right. \nonumber\\
& + & \left(-\frac{273811877}{1088640} + \frac{451}{48}\pi^2 - \frac{11831}{315}\right)\eta 
\nonumber\\
& + &\left. \frac{541}{896}\eta^2 - \frac{5605}{2592}\eta^3 -\frac{856}{105}\ln[16(M\omega)^{2/3}]\right\} (M\omega)^2 \nonumber\\
& + &\left. \left(-\frac{4415}{4032} + \frac{717350}{12096}\eta + \frac{182990}{3024}\eta^2\right)\pi(M\omega)^{7/3}\right]\nonumber \\
\label{omegadot}
\eea
where $\hat{\mathbf{L}}_N$ is a unit vector in the direction of orbital angular momentum (and hence the unit normal to the orbital plane of the binary), $\mathbf{S}_{1,2}$ are the spins, $\chi_{1,2} = |\mathbf{S}_{1,2}|/m_{1,2}$ with $m_{1,2}$ the component masses, $\mathbf{\hat{S}}_{1,2} = \mathbf{S}_{1,2}/|\mathbf{S}_{1,2}|$, $M = m_1 + m_2$ is the total mass, and $\eta = m_1 m_2/M^2$ the symmetric mass ratio. $\gamma = 0.577 \ldots$ is the Euler-Mascheroni constant.
The spins and the direction of the orbital angular momentum evolve according to~\cite{Kidder:1995zr,Apostolatos:1994mx}
\bea
\dot{\mathbf{S}}_1 &=& \frac{(M\omega)^2}{2 M} \left[\eta (M\omega)^{-1/3}\left(4+3\frac{m_2}{m_1}\right)\,\hat{\mathbf{L}}_N \right . \nonumber \\
& &\left. + \frac{1}{M^2}\left(\mathbf{S}_2 - 3(\mathbf{S}_2 \cdot \hat{\mathbf{L}}_N)\right)\,\hat{\mathbf{L}}_N \right] \times {\mathbf{S}}_1, \label{S1dot}\\
\dot{\mathbf{S}}_2 &=& \frac{(M\omega)^2}{2 M} \left[\eta (M\omega)^{-1/3}\left(4+3\frac{m_1}{m_2}\right)\,\hat{\mathbf{L}}_N \right . \nonumber \\
& & \left . + \frac{1}{M^2}\left(\mathbf{S}_1 - 3({\mathbf{S}}_1 \cdot \hat{\mathbf{L}}_N)\right)\,\hat{\mathbf{L}}_N \right] \times \mathbf{S}_2, \label{S2dot}\\
\dot{\hat{\mathbf{L}}}_N &=& - \frac{(M\omega)^{1/3}}{\eta M^2}\dot{\mathbf{S}},
\label{Lhatdot}
\eea 
where $\mathbf{S} = \mathbf{S}_1 + \mathbf{S}_2$. 
The dynamics of the binary is governed by the nonlinear, coupled
differential equations (\ref{omegadot})--(\ref{Lhatdot}). It will not be
possible to solve these exactly, but they can easily be treated numerically. 

By numerically evolving $\omega(t)$ one can obtain the orbital phase,
\be
\Phi(t) = \int^t \omega\, dt. 
\label{phase}
\ee
which can be substituted into the usual expressions for the restricted PN
waveform polarizations~\cite{Wahlquist:1987rx}. In the case of
spinning binaries, we need to take into account the time-dependence of the
amplitudes through the inclination of the orbit with respect to the observer.
The plus and cross polarizations of the gravitational wave are given by
\bea
h_+(t) &=& -\left[1 + (\hat{\mathbf{L}}\cdot\hat{\mathbf{n}})^2\right]\,\cos(2\Phi(t)), \nonumber\\
h_\times(t) &=& -2 (\hat{\mathbf{L}}\cdot\hat{\mathbf{n}})\,\sin(2\Phi(t)), 
\label{hpluscross}
\eea 
where the unit vector $\hat{\mathbf{n}}$ points from source to detector. The detector strain is
\be
h(t) = F_+(t) h_+(t) + F_\times(t) h_\times(t),
\label{h(t)}
\ee
where $F_{+,\times}(t) = F_{+,\times}(\alpha,\delta,\Psi(t))$ are detector
antenna factors which depend on the the right ascension and declination of the
source and a time-dependent polarization angle $\Psi(t)$ (see, e.g.,
Ref.~\cite{Apostolatos:1994mx}).

As suggested by Eq.~(\ref{Lhatdot}), the direction of orbital angular momentum
and hence the plane of the inspiral will undergo precession, the effect being
more pronounced for asymmetric systems. It will also be more prominent if the
spins are large, and if they are significantly misaligned with the orbital
angular momentum. The time evolution of spins and angular momentum will affect
the phasing of the waveform through Eqns.~(\ref{omegadot})-(\ref{phase}), and
the precession of the orbital plane will modulate the amplitudes of the wave
polarizations in Eq.~(\ref{hpluscross}).  The waveforms given by
Eqns.~(\ref{omegadot})--(\ref{h(t)}) will be the ``target-signal waveforms" for
testing our template banks.

\section{A detection template family for spinning black hole binaries}
\label{s:DTF}

The frequency-domain phenomenological detection template family proposed in
Ref.~\cite{Buonanno:2002fy} is designed to capture spin-induced amplitude and
frequency modulation in an approximate way.  Specifically, for gravitational-wave
frequencies $f > 0$, the BCV-spin template is
\be
\tilde{h}[t_0,\alpha_j'](f) = e^{2\pi i f t_0} \Theta(f_{\rm{cut}} - f)\,\left[\sum_{j=1}^3(\alpha_j' + i \alpha_{j+3}')\,h_j(f)\right].
\label{DTF}
\ee
Here $t_0$ is the time of arrival, $\Theta(x)$ is the usual Heaviside step
function
and $f_{\rm{cut}}$ is an upper cut-off frequency beyond which the waveform is
unlikely to be close to a true signal (due to breakdown of the adiabatic
approximation to the inspiral regime). The detection statistic will be  maximized analytically over the parameters $\alpha_1', \ldots, \alpha_6'$ in the linear
combination (\ref{DTF}), as well as over $t_0$; these parameters are referred to as \emph{extrinsic} parameters because they do not need to be explicitly searched over. 

The waveforms $h_j(f)$, $j=1, \ldots,
3$ are basis templates, which take the form 
\be
\tilde{h}_j(f) = \mathcal{A}_j(f)\,e^{i\Phi_{\rm{NM}}(f)},
\label{basistemplates} 
\ee   
where 
\bea
\mathcal{A}_1(f) &=& f^{-7/6},\nn\\
\mathcal{A}_2(f) &=& f^{-7/6}\,\cos(\beta f^{-2/3}),\nn\\
\mathcal{A}_3(f) &=& f^{-7/6}\,\sin(\beta f^{-2/3}),
\eea
and $\beta$ captures the effect of spin-induced amplitude modulation. The
(non-modulated) phase $\Phi_{\rm{NM}}(f)$ takes the form\footnote{What is
called $\psi_3$ here was denoted $\psi_{\frac{3}{2}}$
in~\cite{Buonanno:2002fy}.}
\be
\Phi_{\rm{NM}}(f) = f^{-5/3} (\psi_0 + \psi_3 f).
\ee
It will not be possible to analytically maximize the detection statistic over the parameters $\psi_0$, $\psi_3$, and $\beta$, and these must be explicitly searched over using a bank of templates; they
are referred to as \emph{intrinsic} parameters.

It will often be useful to approximately identify the intrinsic parameters
with the physical masses and spins of a compact binary. By relating
$\psi_0$ and $\psi_3$ to the 0PN and 1.5PN phase
coefficients~\cite{Arun:2004hn}, one has the correspondences
\be
\psi_0 \,\longleftrightarrow \,\frac{3}{128\eta}(\pi M)^{-5/3},\,\, 
\psi_3 \,\longleftrightarrow \, -\frac{3\pi}{8\eta}(\pi M)^{-2/3}.
\label{psicorrespondence}
\ee  
Similarly, the parameter $\beta$ can be related to the rate 
of precession by~\cite{Apostolatos:1994mx}
\be
\beta \,\,\,\longleftrightarrow\,\,\, 256\,\mbox{Hz}^{2/3} \left(1+\frac{3m_2}{4m_1}\right)\frac{m_1}{m_2}\left(\frac{M_\odot}{M}\right)^{2/3}\frac{|\mathbf{S}_1|}{m_1^2}. 
\label{betacorrespondence}
\ee
We stress that these mappings are only approximate, and for a given physical
signal, the detection template that matches best may correspond to values of
$(\psi_0, \psi_3, \beta)$ that differ significantly from the ones suggested by
the identifications above. 
 
The identifications (\ref{psicorrespondence}) allow us to make a choice for
$f_{\rm{cut}}$. In the limit where one component mass goes to zero while total
mass $M$ remains fixed, and assuming zero spins, the frequency of last stable
orbit (LSO) of a test mass in the Schwarzschild spacetime is given by
$f_{\rm{LSO}}(M) = (6^{3/2}\pi M)^{-1}$.  For simplicity we set $f_{\rm{cut}}
= f_{\rm{LSO}}(M)$, where $M = -\psi_3/(16\pi^2 \psi_0)$ is computed from the
correspondence (\ref{psicorrespondence}).

Next one constructs an orthonormal basis from the basis templates
(\ref{basistemplates}) with respect to the usual inner product for waveforms
$a$, $b$ on the template manifold given by
\be
\langle a , b \rangle = 4\Re \int_{f_{\rm s}}^{f_{\rm{cut}}} \frac{\tilde{a}(f)\tilde{b}^\ast(f)}{S_n(f)}df, 
\ee
where tilde denotes a quantity computed directly in the frequency domain (as
in the case of the BCV-spin templates) or the Fourier transform of a
time-domain quantity (such as the waveforms $h(t)$ given in Eq.~(\ref{h(t)})).
$S_n(f)$ is the one-sided power spectral density (PSD) of the detector data,
and $f_{\rm s}$ is some lower cut-off frequency associated with the detector; 
in the case of initial LIGO one sets $f_{\rm s} = 40\,\mbox{Hz}$.
The orthonormalization of the basis templates can be effected using the
Gram-Schmidt procedure as in~\cite{Abbott:2007ai}. In addition one demands
that the templates themselves are normalized (denoted by $\hat{h}$): $\langle \hat{h}, \hat{h} \rangle = 1$. This leads to the
requirement 
\be
\sum_{j=1}^6 \alpha_j^2 = 1,
\label{constraint}
\ee
where the $\alpha_j$, $j = 1, \ldots, 6$, are the coefficients of $\hat{h}$
when expressed into the orthonormal basis of templates resulting from the
Gram-Schmidt procedure.

Finally, the signal-to-noise ratio (SNR), which is used as the BCV-spin
detection statistic, is given by 
\be
\rho = \left[\max_{t_0,\alpha_j} \langle s , \hat{h}[t_0,\alpha_j] \rangle\right]^{1/2},
\label{SNR}
\ee
where $s$ represents the detector data stream, and the maximization over the
$\alpha_j$ is subject to the constraint (\ref{constraint}).

\section{Template banks for spinning binaries}
\label{s:Banks}

The template waveforms $h$ may not exactly model gravitational-wave signals $s$.
The loss in SNR due to differences between the template and signal waveforms is
quantified by the \emph{fitting factor}
$\mathcal{F}$~\cite{Apostolatos:1994mx}. If $s$ is a signal waveform and $h$ a
template waveform, then
\be
\mathcal{F} \equiv \max_{\hat{h}} \langle \hat{s}, \hat{h} \rangle,
\ee
where hat denotes normalization: $\langle \hat{s}, \hat{s} \rangle = \langle 
\hat{h}, \hat{h} \rangle = 1$.   
$1-\mathcal{F}$ is the fractional loss in SNR resulting from the use of 
sub-optimal template waveforms rather than the true signal waveforms.
Since we do
not \emph{a priori} know the intrinsic parameters of any gravitational-wave
signals we may detect, we decide on a target signal space and construct a
discrete bank of templates to search for signals in this space. If $\hat{h}_b$
is a normalized template waveform in the discrete bank and $\hat{h}$ is a
normalized waveform from the space used to construct the bank, then the
minimum match $\mathcal{M}$ of the bank is defined to be~\cite{Owen:1995tm} 
\be
\mathcal{M} \equiv \min_{\hat{h}} \left( \max_{\hat{h}_b \in \text{bank}} \langle \hat{h}, \hat{h}_b \rangle
\right)
\label{MM}
\ee
A typical choice for the minimum match in gravitational wave searches is
$\mathcal{M} = 0.97$. When measuring the performance of a template bank, we
are interested in the \emph{effective fitting factor}
$\bar{\mathcal{F}}$ given by~\cite{Lindblom:2008cm}
\be
\bar{\mathcal{F}} = \max_{\hat{h}_b \in \text{bank}} \langle \hat{s}. \hat{h}_b \rangle
\label{BestMatch}
\ee
If the signal waveforms are identical to those used to construct the bank,
then the effective fitting factor will be bounded below by the minimum
match. In practice, the true gravitational-wave signals will differ from the
templates used to construct the bank, so the effective fitting factor may be
smaller than the minimum match. The larger the effective fitting factor, the
better the bank is at capturing the target signals.

If the parameters $\vec{\lambda}$ between two (normalized) templates differ by a 
small amount $\Delta \vec{\lambda}$, the loss in SNR can be related to a 
distance defined by a metric $g_{ij}$ given by
\be
\left\langle\hat{h}(\vec{\lambda}), \hat{h}(\vec{\lambda} + \Delta \vec{\lambda})\right\rangle \simeq 
1 - g_{ij} \Delta \lambda^i \Delta \lambda^j,
\label{metric}
\ee
where
\be
g_{ij} = 
\frac{1}{2}\left\langle \frac{\partial \hat{h}}{\partial \lambda^i},
\frac{\partial \hat{h}}{\partial \lambda^j}  \right\rangle.
\ee
The standard method of constructing a template bank then consists of computing
this metric in the intrinsic parameter space of a waveform family and using
it to place templates such that the distance between any template waveform and
the nearest template in the bank is greater than the desired minimum match
$\mathcal{M}$.  In
searches for non-spinning binaries, the intrinsic parameters of the templates
are the just component masses $(m_1,m_2)$ of the binary. In practice, we
re-parameterize the waveforms using the chirp times
$(\tau_0(m_1,m_2),\tau_3(m_1,m_2))$~\cite{Owen:1998dk}. With respect to these variables the metric is 
almost Euclidean, and so template placement using the metric $g_{ij}$ becomes
a straightforward two-dimensional hexagonal packing problem~\cite{Babak:2006ty}.

As described in Sec.~\ref{s:intro}, a search for gravitational waves using the
BCV-spin templates has been performed in S3 LIGO data~\cite{Abbott:2007ai}. The metric used in
that analysis was computed using the ``strong modulation approximation'' where
one assumes that the binary precesses many times while emitting in the most sensitive part of the detector's band. This allows one to treat the basis
templates of Eq.~(\ref{basistemplates}) as orthonormal, simplifying the calculation of the metric.  However, the resulting template banks were only appropriate for fairly low-mass, asymmetric systems.  We now present an improved algorithm for constructing a metric in which the assumptions of~\cite{Abbott:2007ai} are dropped. In our case, the parameters of the waveform are
\be
\vec{\lambda} = (t_0,\alpha_1,\ldots,\alpha_6,\psi_0,\psi_3,\beta).
\ee
The detection statistic can be maximized over the extrinsic parameters
$t_0$ and $\alpha_1, \ldots, \alpha_6$, which, as shown in \cite{Pan:2003qt}, leads to a projected metric $g^{\rm{proj}}_{ij}$ which only measures distances in the $(\psi_0,\psi_3,\beta)$ directions. However, the \emph{components} of $g^{\rm{proj}}_{ij}$ will still depend on the $\alpha_j$. This residual dependence on extrinsic parameters can be removed as follows:
\begin{enumerate}
\item Introduce some fiducial distance $\Delta s_0$;
\item Specify a large number of unit vectors (in the coordinate sense) $\hat{n}$ in $(\psi_0,\psi_3,\beta)$ space;
\item For each $\hat{n}$, numerically maximize the \emph{metric} length $\Delta s_{\hat{n}}$ computed from $g^{\rm{proj}}_{ij}$, over values of the $\alpha_j$ consistent with the constraint (\ref{constraint}); i.e., 
\be
\Delta s_{\hat{n}}^2 = \max_{\sum_k\alpha_k^2=1} g^{\rm{proj}}_{ij}(\alpha_m, \beta) \hat{n}^i \hat{n}^j;
\ee
\item Rescale each vector $\hat{n}$ by defining a new vector $\bar{u} = (\Delta s_0/\Delta s_{\hat{n}})\,\hat{n}$;
\item Fit an ellipsoid in parameter space to the vectors $\bar{u}$; 
\item Define an ``effective" metric $g^{\rm{eff}}_{ij}$ by requiring that any point on the ellipsoid is at effective metric distance $\Delta s_0$ from the template we started with.  
\end{enumerate}  
Note that this construction is independent of the fiducial length scale $\Delta s_0$. In what follows, 
$g^{\rm{eff}}_{ij}$ is the metric we will use to satisfy the criterion (\ref{MM}) through the relationship (\ref{metric}). 
An property of $g^{\rm{eff}}_{ij}$ is that it is essentially independent of 
$\psi_0$ and $\psi_3$ and only has a 
weak dependence on $\beta$.

It is important to note that given a short straight line segment in coordinate
space with coordinate length $\Delta\vec{\lambda}$, by construction
$g^{\rm{eff}}_{ij}$ associates almost the largest possible metric length to it
consistent with the family of metrics $g^{\rm{proj}}_{ij}(\alpha_j,\beta)$
parametrized by the $\alpha_j$. When generating template banks, in practice
one specifies a minimum match which will then be used together with the metric
to determine the spacing of templates. Since $g^{\rm{eff}}_{ij}$ is too
conservative in assigning lengths, neighboring templates will tend to have a larger
match than needed, and the true minimum match defined by (\ref{MM}) will
always be significantly larger than what was originally intended. As we shall see below, 
setting an \emph{a priori} value of 
$\mathcal{M} = 0.8$ will be more than enough for a bank to obtain high overlaps 
($\gtrsim 0.9$) with target waveforms.

We would like to capture signals from binaries whose component masses are in
the interval $[1,35]M_\odot$, with total masses $M \leq 35\,M_\odot$. We do
not need to worry about capturing BNS signals, since spin does not have a
significant effect on waveforms from those sources. However, our template bank
should have good overlap with NS-BH and BBH signals.  Taking neutron star
masses to lie between $1\,M_\odot$ and $3\,M_\odot$ and black hole masses to
be larger than $3\,M_\odot$, we impose $M \geq 4\,M_\odot$.  To capture these
signals, we want an appropriately chosen bounding box in
$(\psi_0,\psi_3,\beta)$ within which to place templates.
Such a box can be specified using the correspondences
(\ref{psicorrespondence})--(\ref{betacorrespondence}). The suggested intervals for $(\psi_0,\psi_3)$ are
then roughly \bea
\psi_0 \in [8 \times 10^3, 5 \times 10^5]\,\mbox{Hz}^{5/3},\nonumber\\
\psi_3 \in [-3 \times 10^3, 10]\,\mbox{Hz}^{2/3},
\label{psi0psi3ranges}
\eea
where the upper bound for $\psi_3$ has been chosen generously. 
As to $\beta$, the correspondence (\ref{betacorrespondence}) suggests that $\beta \lesssim 150\,\mbox{Hz}^{2/3}$ 
should suffice, but to have good matches with a variety of physical signals, here too it turned out to be better 
to have a larger upper bound:
\be
\beta \in [1,4 \times 10^2]\,\mbox{Hz}^{2/3}.
\label{betarange}
\ee
We now present two methods for constructing template banks for BCV-spin
templates which cover this space.

\subsection{Square-hexagonal template bank}

The metric $g^{\rm{eff}}_{ij}$ depends only on $\beta$, so it is natural to
first define layers of constant $\beta$, with a spacing determined by the
minimum match. Within each of the two-dimensional layers one can then lay out
templates in a hexagonal pattern, which is the optimal placement in two
dimensions. We will refer to this kind of placement as
\emph{square-hexagonal}.  The construction of this bank is analogous to that
described in Ref.~\cite{Cokelaer:2007kx} which was used to construct template
banks for search for binary black holes in data from the third and fourth LIGO
science runs~\cite{Abbott:2007xi} using non-spinning phenomenological
templates~\cite{Buonanno:2002ft}.  For the BCV-spin templates, we have a
3-metric, which in each $\beta$ layer is diagonalized by going to a new set of
coordinates $(\psi_0',\psi_3',\beta')$, where $\beta'=\beta$. After that a
hexagonal placement in $(\psi_0',\psi_3')$ can be performed as
in~\cite{Cokelaer:2007kx}.  As explained above, the metric is overly
conservative in specifying distances between templates, and setting an \emph{a
priori} minimum match of $\mathcal{M} =0.8$ will suffice to obtain high
matches with target waveforms.

\subsection{Stochastic template bank}
\label{ss:Stochastic}

We now consider a different bank placement for BCV-spin, which we hope will reduce the overcoverage of the parameter space that is unavoidable with the square-hexagonal placement method defined above. This will lead to a smaller number of templates but will yield the same or better matches with target waveforms, and similar efficiencies. This template bank is created by the placement of a large number of randomly distributed templates, followed by a ``pruning'' stage in which unnecessary templates are discarded. This method is described in~\cite{Allen:2009} and summarized below. Other, similar methods for creating stochastic template banks were proposed in~\cite{Babak:2008rb} and~\cite{Messenger:2008ta}.

The stochastic placement algorithm we wish to use for BCV-spin is very simple. We begin by generating a very large number
of points in the parameter space, far more than would be needed to fill the space. We then iteratively cycle through these points, retaining a point only if it is not closer than some predefined metric distance $\Delta$ to the points retained in previous iterations. The remaining points form our stochastically generated bank. Tests have shown~\cite{Allen:2009} that one should begin with at least $N^{1.5}$ points, where $N$ would be the number of templates remaining after filtering, to have a good coverage of the parameter space after ``pruning''. 

In testing this algorithm against lattice placement algorithms it was found~\cite{Allen:2009} that in a 2-dimensional Cartesian space the stochastic algorithm produced a template bank with 1.5 times the number of templates that a square lattice algorithm would have generated. However, in the case of a 2-dimensional non-spinning (non-Cartesian) SPA bank (as described above) the stochastic algorithm was found to place $\sim 10$\% less templates than the square lattice algorithm and only $\sim 25$\% more templates than the hexagonal lattice placement, while achieving a similar degree of coverage. We emphasize here that this stochastic placement algorithm would be of most use in parameter spaces with more than 2 dimensions, where lattice placement becomes significantly sub-obtimal.

For the specific case of BCV-spin the templates are sprinkled randomly over a rectangular box in $(\psi_0,\psi_3,\beta)$ space using the same bounding box as in the previous subsection.
An estimate for the number of templates that will be needed is provided by the invariant volume of the box, divided by the volume taken up by an individual template:
\be
\mathcal{N} = \frac{\int_{\rm box} \sqrt{\mbox{det}(g^{\rm{eff}}_{ij})}\,d\psi_0 d\psi_3 d\beta}{(1-\mathcal{M})^{3/2}}.
\ee 

Once again it will suffice to set an \emph{a priori} minimum match $\mathcal{M} = 0.8$ (i.e., setting the $\Delta$ defined above to 0.2). Given the box in parameter space specified by (\ref{psi0psi3ranges},\ref{betarange}), the number of sprinkled templates should then be about 500,000. When using a larger number of initial templates, we find that the final number of templates after pruning does not change significantly.
With the Initial LIGO design PSD, the number of templates for stochastic BCV-spin banks with $\mathcal{M}=0.8$ is about 8000; SPA banks with $\mathcal{M}=0.95$ have $\sim 12,000$ templates, and for BCV-spin with square-hexagonal placement and $\mathcal{M}=0.8$ more than $16,000$ templates are obtained (see Table \ref{banktable}).

\section{Comparison of BCV-spin banks with spinning PN signals}
\label{s:Matches}

We now study the performance of our banks against the target
waveforms of Sec.~\ref{s:TargetWaveforms}. In particular, for a variety of
target waveforms $s$ corresponding to different masses and initial spins, we
compute the effective fitting factor $\bar{\mathcal{F}}$ of the bank for the
target waveforms, as given by Eq.~(\ref{BestMatch}).

\begin{figure}[htbp!]
\centering
\includegraphics[scale=0.35,angle=0]{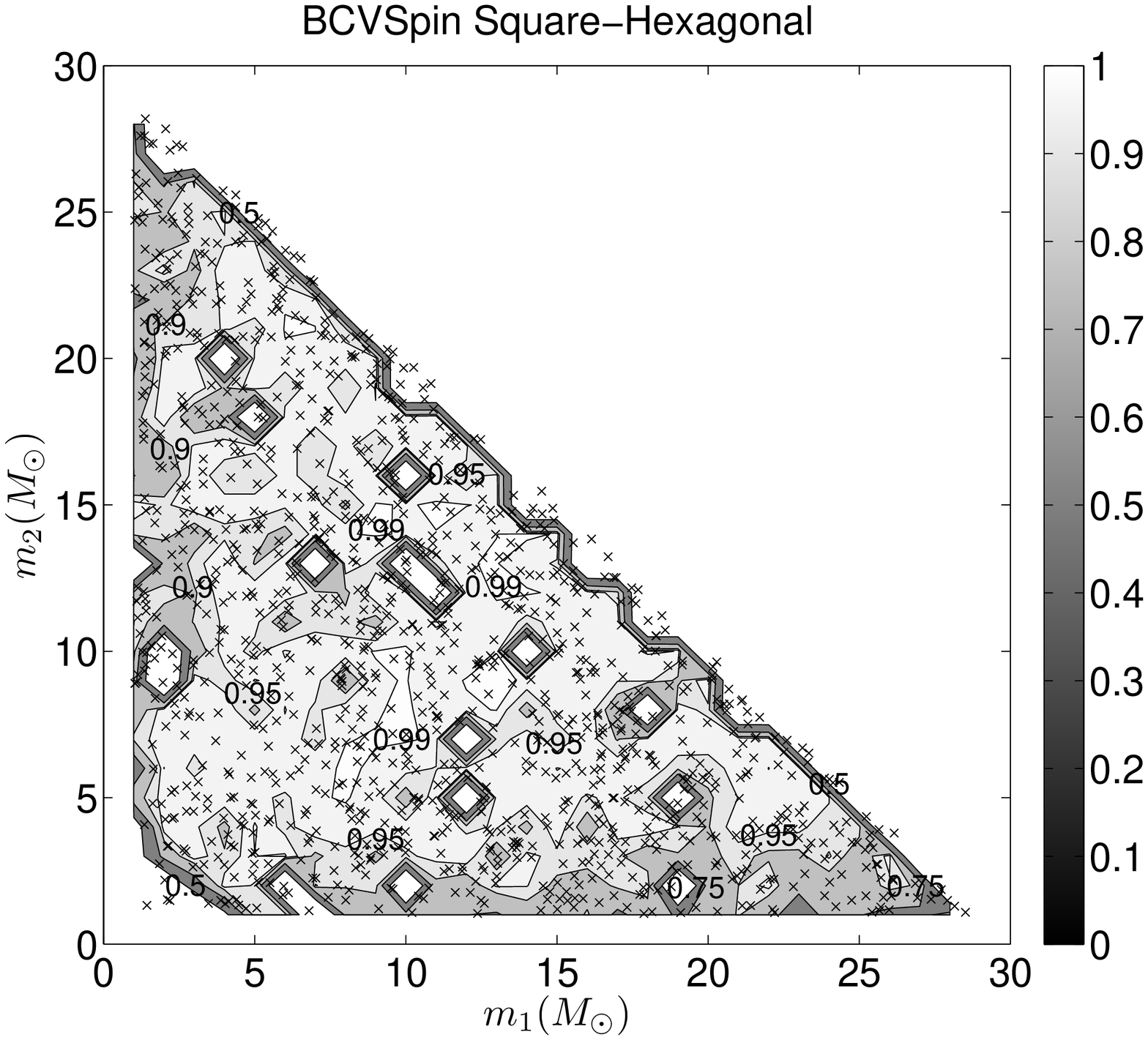}
\includegraphics[scale=0.35,angle=0]{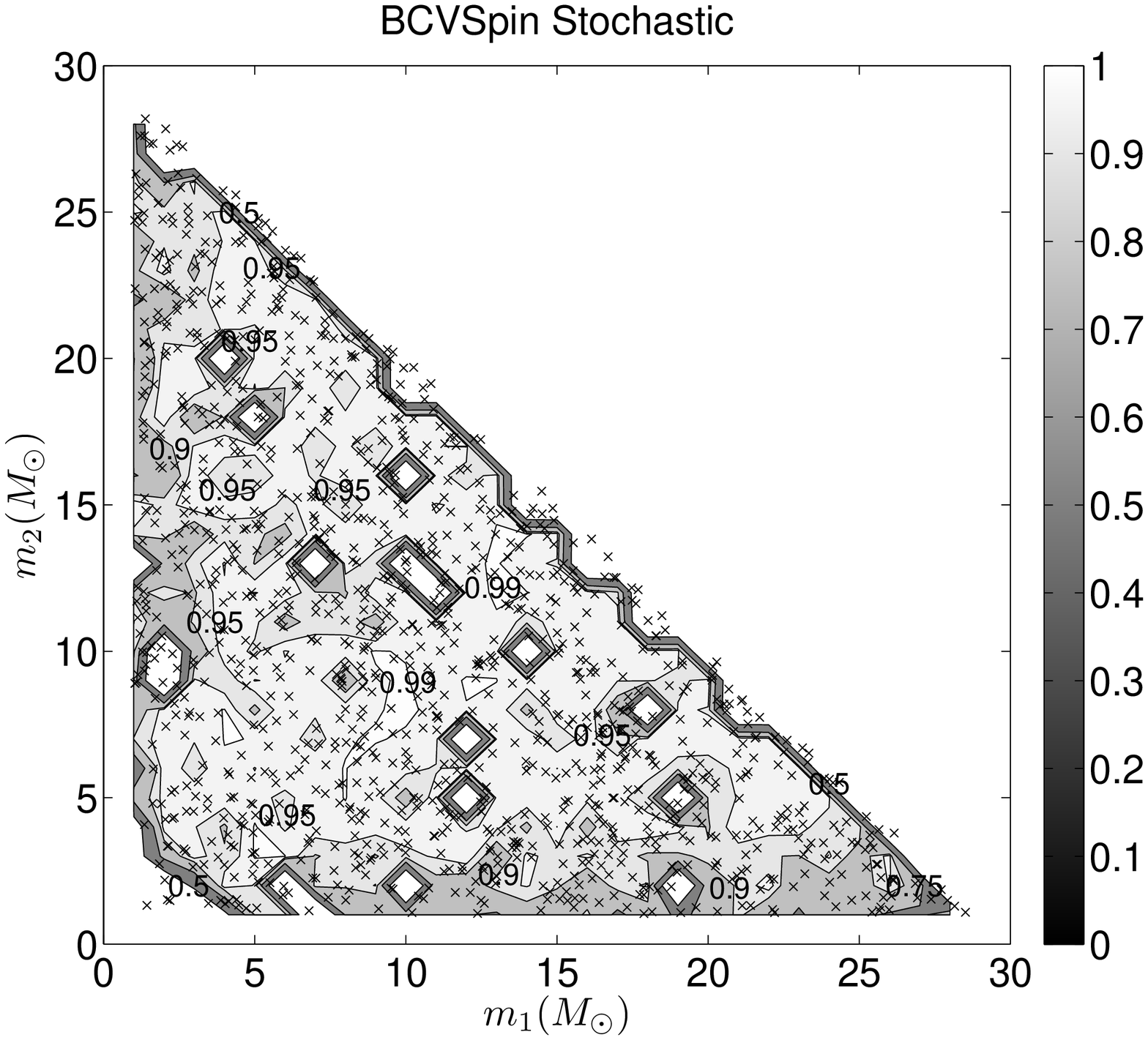}
\includegraphics[scale=0.35,angle=0]{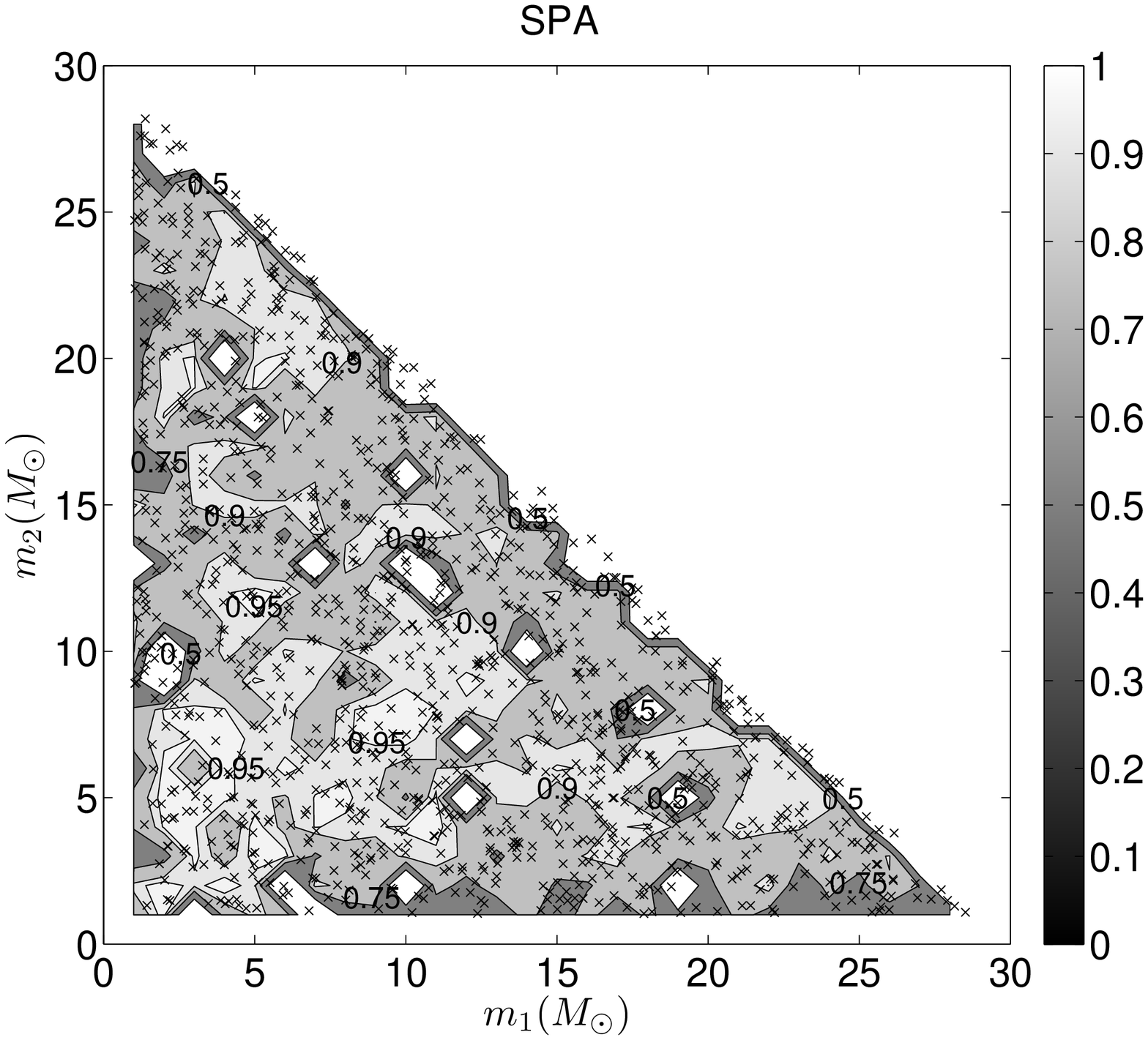}
\caption{Effective fitting factors of 1124 target waveforms with templates in a square-hexagonal BCV-spin bank (top), a stochastic BCV-spin bank (middle), and an SPA bank using the Initial LIGO design PSD. The black crosses indicate the component masses for the target waveforms; spins have random orientations, and $0.7 \leq |\mathbf{S}_{1,2}/m_{1,2}^2| \leq 1$. The color coding gives the effective fitting factor of a target waveform over the template bank. There is no discernable difference between the performances of the square-hexagonal and stochastic BCV-spin banks, but both do notably better than SPA.}
\label{f:Matches}
\end{figure}  

Fig.~\ref{f:Matches} compares the effective fitting factor of templates in square-hexagonal
and stochastic BCV-spin banks with those of a non-spinning SPA bank.
There is no discernable difference between square-hexagonal and stochastic
placements, but both differ significantly from the non-spinning SPA bank. As
one would expect, the difference is largest for binaries with large mass
ratios, although there is improvement also for a variety of other target
waveform masses.  Depending on masses and spins, for the same target waveform
$s$, the difference in $\bar{\mathcal{F}}$ can be more than $25\%$.  The medians
and means for the effective fitting factors are summarized in Table \ref{banktable}.  We find
that BCV-spin with stochastic bank placement has marginally better effective fitting factors than the square-hexagonal bank, and both BCV-spin banks have noticably higher effective fitting factors
than the non-spinning bank. Given the small difference between the stochastic
and square-hexagonal BCV-spin banks we will subsequently only consider
differences between the stochastic BCV-spin bank and the non-spinning SPA bank.

\begin{table*}[t]
\caption{Number of templates for the SPA, square-hexagonal, and BCV-spin banks with Initial LIGO design PSD, and median and mean effective fitting factors $\bar{\mathcal{F}}$ of the banks with target waveforms as in Fig.~\ref{f:Matches}}
\label{banktable}
\centering{}
\begin{tabular}{c|c|c|c|c|c|c}
\hline
\hline
Template & Bank-placement & SNR threshold & Minimum match $\mathcal{M}$ & Number of templates & $\bar{\mathcal{F}}_\mathrm{median}$& $\bar{\mathcal{F}}_\mathrm{mean}$\\
\hline
SPA     & SPA              & 5.5 & 0.95 & 11832  & 0.89 & 0.86 \\
BCV-spin & Square-hexagonal & 8   & 0.8  & 16431  & 0.96 & 0.92 \\
BCV-spin & Stochastic       & 8   & 0.8  & 7913   & 0.96 & 0.93 \\
\hline
\end{tabular}
\end{table*}

\section{Search performance of BCV-spin template banks}
\label{s:False alarms}

The effective fitting factor of a target waveform over a template bank as defined in
Eq.~(\ref{BestMatch}) indicates how similar the templates are to physical
signals, but when searching for gravitational-wave signals in real detector
data, other factors also come into play. The effective fitting factor of a template bank gives us
a measure of how the signal-to-noise ratio is reduced by not filtering with
the true signal waveform, but to detect a signal we must be able to
distingusih it from background noise in the detector. To determine the overall
performance of a template bank, we have to consider both the effective fitting
factor and the
\emph{false alarm rate} of the bank, i.e., the response of the filters to
noise (both Gaussian and transient) in the detector. Once we establish the
false alarm rate of a search, we measure the performance of a bank by its
\emph{efficiency}, i.e., the bank's ability to find simulated target waveforms
injected in the noise at a given false alarm rate. We will establish the false alarm rates and efficiencies of BCV-spin and SPA banks by means of the data-analysis pipeline used in searches by the LIGO Scientific Collaboration (LSC) for inspiral signals~\cite{Abbott:2007xi,Abbott:2007ai,Abbott:2009tt}, which is available in the LSC Algorithm Library~\cite{lal}. More details on this pipeline can be found in Ref.~\cite{LIGOS3S4Tuning}.

\subsection{False alarm rates}

The BCV-spin detection statistic (\ref{SNR}) involves maximization over six
parameters ($t_0$, and the $\alpha_j$ with the constraint $\sum_j \alpha_j^2 =
1$), to be compared with only two for SPA.  It should also be noted that the
BCV-spin detection template family consists of waveforms that are only
approximate.  As we shall see below, the larger number of degrees of freedom
will make the BCV-spin banks more prone to detecting instrumental noise
transients with high SNR.  Both for SPA and BCV-spin one needs to set an SNR
threshold below which no candidate events are accepted, and the higher false
alarm rate with BCV-spin will necessitate setting a higher threshold. 

The pipeline used to search for gravitational-wave signals in the LIGO
detectors demands that candidate events be coincident in two or more
detectors~\cite{LIGOS3S4Tuning}. If the noise sources in our detectors are
uncorrelated (as in the case of the two $4$~km LIGO detectors), we can measure
the false alarm (or background) rate of this pipeline by time-shifting the
detector data by more than the gravitational-wave travel time between the
detectors ($11$~ms) and looking for coincident triggers; such triggers will be
due to accidental coincidence of noise alone. We can repeat this with time
steps of, say, $10$~s, and count the number of coincident triggers in
each of the time-shifts to obtain a good estimate of the false alarm rate.

Before triggers are compared between detectors, they are clustered together,
keeping only the trigger with the loudest SNR within a certain time window (in
our case $4000$~ms). Next, various methods can be used for declaring two
clustered triggers to be coincident across detectors. Usually one demands not
only coincidence in time, but also that the parameters of the template that
gave the loudest SNR be similar in the different detectors. The simplest way
of implementing this is the so-called box-coincidence method, whereby two
triggers are considered coincident if they occured within a certain time from
each other (say, 100 ms), and the associated templates have parameters that
differ only within certain tolerances~\cite{LIGOS3S4Tuning}. In the case of
BCV-spin, these were chosen as $\Delta\psi_0 = 40000\,\mbox{Hz}^{5/3}$ and
$\Delta\psi_3 = 600\,\mbox{Hz}^{2/3}$, with no restrictions on differences in
$\beta$~\cite{Abbott:2007ai}. 

More recently, a more sophisticated technique was developed which has the
potential to dramatically reduce the false alarm rate~\cite{Robinson:2008un}.
In this method, the covariances between the signal parameters are used to
define an error ellipsoid in parameter space around the triggers, and triggers
in different detectors are considered coincident if their associated
ellipsoids overlap. In the case of SPA banks, the size of the ellipsoids will
depend strongly on the region of parameter space the triggers occur in.
Generally they will be smaller for triggers associated with smaller masses, as
waveforms will then spend more time in the detector band and errors will be
smaller. This leads to a dramatic reduction in the number of spurious
coincident triggers. By contrast, the box-coincidence method described above
uses the same parameter windows anywhere in parameter space. 

The ellipsoid coincidence method has been successfully implemented for SPA
banks. The technique is well-suited for banks where the templates are
simplified versions of target waveforms, so that one can assume template
waveforms to be reasonably close to physical signals. It would be possible in
principle to implement such a method also for the (phenomenological) BCV-spin
banks. However, in this case the metric $g^{\rm{eff}}_{ij}$ is basically independent of
$\psi_0$ and $\psi_3$, the parameters that are most closely related to the
masses. Hence, for a given value of $\beta$, the associated ellipsoids would
not differ in size across $(\psi_0,\psi_3)$ space, and only their orientations
would differ with $\beta$. This way, no great improvements can be expected in
terms of reducing the false alarm rate.  

\begin{table*}[t]
\caption{Average number of triggers per time-shifts ($\langle N \rangle$), and variances thereof ($\sigma = \langle N^2 \rangle^{1/2}$), for SPA with $\mathcal{M} = 0.95$, and BCV-spin with stochastic placement and $\mathcal{M} = 0.8$. Note that the 1-$\sigma$ intervals overlap, so that the false alarm rates are comparable. Next to these we list distances at which the efficiencies are 50\%, 75\%, and 90\%. See~\cite{Abbott:2009tt} for histograms of trigger numbers and efficiency plots.}
\label{time-shifts}
\centering{}
\begin{tabular}{c|c|c||c|c|c|c}
\hline
\hline
Bank & $\langle N \rangle$ & $\sigma$ & $D_{50\%}$ (Mpc) & $D_{75\%}$ (Mpc) & $D_{90\%}$ (Mpc) \\
\hline
SPA     & 97.3 & 8.7 &  40.1 & 33.9 & 15.9 \\
BCV-spin & 85.4 & 8.4 &  34.6 & 17.5 & 14.5 \\
\hline
\end{tabular}
\end{table*}

Table.~\ref{time-shifts} shows the average number and variance of coincident
triggers between the $4$~km LIGO Hanford and Livingston detectors for
time-shifts within $\sim 9$ days of data from the fifth LIGO science
run~\cite{Abbott:2007kva}. The SNR threshold for SPA is 5.5, while for
BCV-spin it is 8. With these thresholds, SPA and BCV-spin banks have
approximately the same false alarm rates.

\subsection{Efficiencies}

We are now in a position to compare the efficiencies of SPA and BCV-spin
banks. Given a large number of target waveforms injected in the data, the
efficiency is the ratio of the number of found injections to the total number
of injections made. For our purposes, an injection is considered found if it
had an SNR above the chosen threshold with at least one template in the bank,
within a certain time interval around the time when the injection was actually
made. In the case of SPA, the width of this interval can be chosen to be 40
ms. BCV-spin templates, being phenomenological, turn out to have a larger
timing inaccuracy, and an interval of 100 ms was found to be more appropriate.
This had already been noticed in~\cite{Abbott:2007ai}; presumably the larger
timing uncertainties of BCV-spin are related to its unphysical phasing
(essentially, missing PN terms) as it is predominantly the phasing
which affects timing errors.

It is important that efficiencies be compared \emph{for the same false alarm
rate}. And indeed, as we have just seen, SPA and BCV-spin have essentially the
same background rates if the SNR thresholds are set at 5.5 and 8,
respectively. 

We made 1124 injections distributed logarithmically in distance between 1 Mpc
and 50 Mpc, with component masses randomly chosen between $1\,M_\odot$ and
$30\,M_\odot$, spin magnitudes $|\mathbf{S}_i|/m_i^2$, $i=1,2$ between 0.7 and
1, and arbitrary directions for the initial spin vectors. The efficiency of
SPA then came out to be 0.93, versus 0.89 for BCV-spin. These results have been
summarized in Appendix I of~\cite{Abbott:2009tt}; here we have provided a
detailed account of how they were obtained. We refer to the latter paper for
plots of efficiency against distance; see Table \ref{time-shifts} for the
distances at which the efficiencies are 50\%, 75\%, and 90\%, both for SPA and
stochastic BCV-spin.

We find that despite the fact that BCV-spin banks have higher effective
fitting factors with the
target waveforms than SPA banks, in a more realistic data-analysis comparison
the two waveform families have similar abilities to detect simulated signals.
The detection statistic for BCV-spin involves more degrees of freedom and the
pipeline using BCV-spin waveforms is more sensitive to non-stationary noise
transients in the data. Consequently, at the same false alarm rate the
detection threshold of the BCV-spin bank is higher than the SPA bank,
negating the effect of the improved effective fitting factor of the BCV-spin bank\footnote{The
problem had been anticipated in~\cite{Buonanno:2002fy}; here we have
quantified it using real data.}. Searches for spinning binaries using the
non-spinning bank therefore have approximately the same performance as even
our improved BCV-spin bank.

\section{Conclusions}
\label{s:Conclusions}

Past searches for low mass compact binary inspiral events in LIGO data (with
the exception of~\cite{Abbott:2007ai}) have used waveforms which do not
attempt to model the spin effects, despite the fact that astrophysical black holes may be spinning rapidly.
In this paper we have constructed template banks using the BCV-spin waveform
proposed in~\cite{Buonanno:2002fy}. Though phenomenological, these
waveforms seek to capture the spin-induced amplitude modulation one expects to see in a
physical signal, and have high effective fitting factors with PN waveforms that
include spin. 
We have improved on the search method of~\cite{Abbott:2007ai},
in two ways: (i) we have constructed a bank using the metric outlined in
Ref.~\cite{Buonanno:2005pt}, which is much better suited to the template
family, and (ii) we have explored two new placement algorithms
(square-hexagonal and stochastic). 
We used spinning PN signals to study the effective fitting factors of three different
banks: an SPA bank, a BCV-spin bank with square-hexagonal placement, and a
BCV-spin bank with stochastic placement. We found that the two BCV-spin banks
had a similar performance, but both did markedly better than SPA. 
However, search performance should be judged by detection efficiency at a given false
alarm rate. 
The search pipeline for low-mass compact binaries ($2\,M_\odot \le M \le 35\,
M_\odot$) in data from the fifth LIGO science run used a non-spinning SPA bank
with an SNR threshold of $5.5$. We have demonstrated that to achieve a
comparible false alarm rate with the currently available search pipelines
using BCV-spin templates requires an SNR threshold of $8$ and with this higher
threshold, the detection efficiency of BCV-spin for spinning PN signals becomes similar to that of the non-spinning SPA pipeline. Our findings, presented at
length here and summarized in Ref.~\cite{Abbott:2009tt}, were used to
guide the decision not to repeat the analysis of Ref.~\cite{Abbott:2007ai}
with data from the fifth LIGO science run.

In conclusion, the detection performance of the BCV-spin pipeline is similar
to that of the non-spinning SPA pipeline.  We note, however, that our
comparison is not entirely fair, because the SPA pipeline implements the
metric-based coincidence algorithm of Ref.~\cite{Robinson:2008un} which
dramatically reduces the number of spurious coincident triggers.
In principle such a technique could also be applied for BCV-spin, but since
the metric has essentially no dependence on $(\psi_0,\psi_3)$ and only a weak 
dependence on $\beta$ it is unlikely that implementation of the metric-coincidence
algorithm would improve the sensitivity of the BCV-spin pipeline.  This
justifies the use of non-spinning SPA pipelines rather than BCV-spin pipelines
in LIGO searches. Nevertheless, to search for spinning signals with
non-spinning banks is still sub-optimal, and work is ongoing to improve the
performance of searches for spinning signals using templates determined by
physical (rather than phenomenological) parameters proposed in
Ref.~\cite{Pan:2003qt,Buonanno:2004yd}. In the mean time we
recommend the continued use of non-spinning SPA banks in upcoming searches
until more efficient template families designed to capture spin-modulated
waveforms have been incorporated into a pipeline.

\section*{Acknowlegements}
We would like to thank the members of the LIGO Scientific Collaboration
Compact Binary Coalescence group for many helpful discussions, and in 
particular Michele Vallisneri for a critical reading of the manuscript. 
CVDB, TC, IH, GJ, and BSS are supported by PPARC grant PP/F001096/1. DB is supported
by National Science Foundation grant NSF-0847611. Hideyuki Tagoshi is supported by  KAKENHI, Nos.~16540251 and 20540271. Hirotaka Takahashi is partially supported
by the Uchida Energy Science Promotion Foundation, Sasagawa Scientific
Research Grant and the JSPS Grant-in-Aid for Scientific Research No.~20540260.

\newpage
\bibstyle{prd}
\bibliography{SBBHTemplateBanks}

\end{document}